\documentclass[twocolumn,prl,aps]{revtex4}
\usepackage{epsfig}

\begin{document}
\textheight 226mm

\title{\Large Thermodynamic stabilities of ternary metal borides:
An \textit{ab initio} guide for synthesizing layered superconductors}

\author{Aleksey N. Kolmogorov$^1$, Matteo Calandra$^2$, and Stefano Curtarolo$^3$} 

\affiliation{
$^1$Department of Materials, University of Oxford, Parks Road, Oxford OX1 3PH, United Kingdom \\
$^2$Institut de MinΩralogie et de Physique des Milieux condens\'es %,
case 115, 4 place Jussieu, 75252, Paris cedex 05, France\\
$^3$Department of Mechanical Engineering and Materials Science %,
Duke University, Durham, North Carolina 27708, USA 
}

\date{\today}

\begin{abstract} { 
    Density functional theory calculations have been
    used to identify stable layered Li-$M$-B crystal structure phases
    derived from a recently proposed binary metal-sandwich (MS)
    lithium monoboride superconductor. We show that the MS lithium
    monoboride gains in stability when alloyed with electron-rich
    metal diborides; the resulting ordered Li$_{2(1-x)}M_x$B$_2$
    ternary phases may form under normal synthesis conditions in a
    wide concentration range of $x$ for a number of group-III-V metals
    $M$. In an effort to pre-select compounds with the strongest
    electron-phonon coupling we examine the softening of the in-plane
    boron phonon mode at $\Gamma$ in a large class of metal
    borides. Our results reveal interesting general trends for the
    frequency of the in-plane boron phonon modes as a function of the
    boron-boron bond length and the valence of the metal. One of the
    candidates with a promise to be an MgB$_2$-type superconductor,
    Li$_2$AlB$_4$, has been examined in more detail: according to our
    {\it ab initio} calculations of the phonon dispersion and the
    electron-phonon coupling $\lambda$, the compound should have a
    critical temperature of $\sim4$ K. }
\end{abstract}

\maketitle

%\draft

\renewcommand{\topfraction}{0.85}
\renewcommand{\textfraction}{0.1}
\renewcommand{\floatpagefraction}{0.75}

\subsection{I. Introduction}

Observation and explanation of the superconducting transition in
MgB$_2$ at an unexpectedly high temperature of 39 K
 \cite{origin,band_filling} have stimulated extensive research aimed at
developing related layered phonon-mediated superconductors
 \cite{CaAlSi,TaB2,MB2,NbxB2,LiBC,LixBC,SCGIC1,SCGIC2,SCGIC3,US1,MgC}.
Interesting results have been recently obtained for carbon-based
layered materials: the efforts to adjust the properties of graphite
via intercalation with various metals have led to the discovery of a
CaC$_6$ superconductor with a critical temperature of 11.5 K
 \cite{SCGIC1,SCGIC2} (15.1 K under hydrostatic pressure \cite{SCGIC3}).

Tuning the properties of boron-based superconductors has proven to be
difficult because, despite the existence of over a dozen of stable
metal diborides with the C32 structure \cite{ICSD,PAULING,Oguchi},
only one of them, MgB$_2$, has the hole-doped boron $p\sigma$ states
that couple strongly to the in-plane boron phonon modes
\cite{band_filling}. For comparison, critical temperatures in other
C32-$M$B$_2$ superconductors, such as TaB$_2$ or NbB$_2$
\cite{TaB2,MB2,NbxB2}, do not exceed 10 K because these electron-rich
compounds have fully occupied boron $p\sigma$ bands and thus lack the
important nearly two-dimensional Fermi surfaces. The attempts to raise
$T_c$ in MgB$_2$ via substitutional doping have been unsuccessful for
various reasons: i) electron-doping led to the decrease of the
electron-phonon coupling due mainly to the filling of the $p\sigma$
band \cite{band_filling}; ii) hole-doping or substitution of Mg with
large metals turned out to be thermodynamically unfavorable
\cite{dope_review,MgB2_defects}; and iii) co-doping of MgB$_2$ with
Li-Al Li-C caused the reduction of the $T_c$ because of the concurrent
filling of the $p\sigma$ band and depletion of the $p\pi$ band
\cite{MgAlLiB1}. Recently, Palnichenko {\it et al.}  have reported
interesting results on a possible onset of superconductivity in
MgB$_2$ after thermal treatment in the presence of Rb, Cs, and Ba; the
structure of the resultant materials and the mechanism of the $T_c$
enhancement have yet to be determined \cite{Palnichenko,Chepulskyy}.

Identification and synthesis of new stable compounds will be a
critical step to overcome the limitations of the existing layered
metal borides and to have a chance of obtaining better
superconductors. Using a data-mining approach we have recently found a
previously unknown metal sandwich (MS) crystal structure and
demonstrated that lithium monoboride in this configuration is
marginally stable under ambient conditions and is likely to form under
pressure \cite{US1,US2}. Remarkably, our calculations indicated that,
relative to MgB$_2$, MS-LiB has a higher density of boron $p\sigma$
states at the Fermi level \cite{US1,US2}, a feature long sought for in
MgB$_2$-related materials. Subsequent theoretical studies of the
electron-phonon coupling predicted that the $T_c$ in MS-LiB would be
around 10 K \cite{Mazin,US3}. The relatively low value of $T_c$ has
been attributed to an accidental electronic structure feature present
in the pristine MS-LiB: the crossing of the $p\pi$ bands happens to be
exactly at the Fermi level \cite{Mazin,US3}. After examining the
compound's response to hydrostatic pressure and substitutional doping
we have concluded that the recovery of the $p\pi$ density of states
(DOS) at the Fermi level would be accompanied by an unfortunate drop
of the $p\sigma$ DOS under these conditions \cite{US3}. The question of
whether it could be possible to tune the properties of MS-LiB to
obtain higher $T_c$ remains open.

In this work we expand the search for stable crystal structure phases
to ternary metal borides. Namely, we use density functional theory
calculations to explore the relative stability of different layered
configurations with composition Li$_{2(1-x)}M_x$B$_2$ for a library
of over 30 metals $M$. We show that MS lithium monoboride gains in
stability when alloyed with electron-rich metal diborides; the
resulting ordered Li$_{2(1-x)}M_x$B$_2$ ternary phases may form under
normal synthesis conditions in a large concentration range of $x$ for
a number of group-III-V metals $M$. For several promising candidates
we evaluate the softening of the in-plane boron phonon mode at $\Gamma$ 
in order to identify which of these compounds has large
coupling to the in-plane boron modes (similarly to what happens in
MgB$_2$). For one case, Li$_2$AlB$_4$, we also perform a full
calculation of the critical temperature. We believe that our results
will be of considerable help to experimental groups working on the
development of new boron-based layered superconductors.

The paper is organized as follows. In Sec. II we describe the geometry
of the ternary MS structures. Section III is devoted to the
zero-temperature thermodynamic stability of Li$_{2(1-x)}M_x$B$_2$
phases and to the special case Li$_{2(1-x)}$Al$_x$B$_2$. Electronic
properties of two representatives, Li$_2$AlB$_4$ and Li$_2$TiB$_4$,
are addressed in Sec. IV. In Sec. V we examine the phonon softening in
a large set of metal borides. In Sec. VI we present the {\it ab initio}
superconductivity analysis of Li$_2$AlB$_4$. Conclusions are given in
Sec. VII.

\subsection{II. Li$_{2(1-x)}M_x$B$_2$ metal sandwich structures}

Consideration of ternary Li-$M$-B systems dramatically increases the
number of candidates to be screened because one should check both
different compositions and various populations of metal sites. We
reduce the search space by recalling the binding trends for the
binary MS metal borides demonstrated in our previous
study \cite{US2}. Namely, the MS lithium monoboride has plenty of
available bonding boron $p \sigma$ states and may additionally
stabilize when mixed with electron-rich metals. Because
straightforward substitutional doping generally causes strain and loss
of binding within the metal layers \cite{MgB2_defects, US4} we separate
regions with different metal species by boron layers and, thus, avoid
putting different metals in direct contact. This idea follows a
natural segregation tendency in layered borides, which has been
observed, for instance, in MgB$_2$ under heavy Al doping: the
resulting compound was a C32 superstructure with alternating Mg and Al
layers \cite{MgAlB4}.

% here is another one on MgAlB4: PHYSICAL REVIEW B 71, 174506 (2005)

\begin{figure}[t]
 \begin{center}
%   \vspace{-10mm}
   \centerline{\epsfig{file=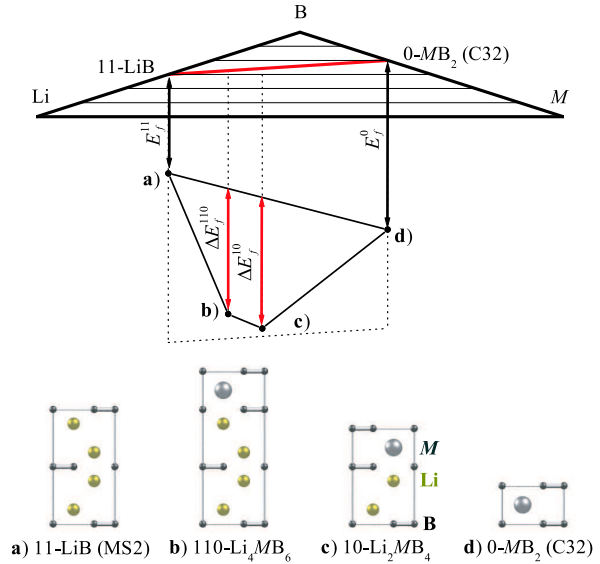,width=80mm,clip=}}
%  \vspace{-5mm}
   \caption{ \small (Color online). Structure of the proposed ternary
    metal sandwich phases and their location on a schematic ternary
    (Li-$M$-B) phase diagram. Formation energies are shown along
    the vertical axis. The relative stability is calculated with
    respect to 11-LiB (MS2) and 0-MB$_2$ (C32) (notation explained in
    the text).}
 \label{fig1}
  \end{center}
% \vspace{10mm}
\end{figure}
%\vspace{-5mm}
%
Our candidate segregated superstructures are built by combining MS-LiB
and C32-$M$B$_2$ unit cells. Since there are numerous possibilities to
stack metal and boron layers, it is convenient to adopt the following
compact notation: MS1-LiB and C32-$M$B$_2$ basic units will be denoted
as ``1''- and ``0''-unit, respectively.  In this way one can uniquely
specify the structure with a string of ones and zeros. Note that the
presence of one ``1''-unit results in a stacking shift along the
$c$-axis which makes the unit cell rhombohedral  \cite{US1,US2}. For
convenience, in candidate phases containing an even number of
``1''-units we choose to reflect every second ``1''-unit in the
$(x,y)$-plane to obtain a hexagonal unit cell. This choice is expected
to not affect our results, since we have found no appreciable
differences in the properties of the rhombohedral and hexagonal
allotropes of MS-LiB  \cite{US1}. Examples of two superstructures with
small unit cells are 10-Li$_2$$M$B$_4$ [Fig. \ref{fig1}(c)] and
110-Li$_4$$M$B$_6$ [Fig. \ref{fig1}(b)], the latter being the
superposition of MS2-LiB and C32-$M$B$_2$.

Figure \ref{fig1} shows that such compounds are located along the red
line on the ternary phase diagram. Note that MS-LiB is marginally
stable (the known competing phase LiB$_y$ has a completely different
structure  \cite{US2}). This implies that if (i) the C32-MB$_2$ exists
and (ii) the relative stability of the MS ternary phases along this
line is negative, then one would have a better chance of synthesizing
the MS ternary compounds from elemental materials rather than the two
binaries MS2-LiB and $M$B$_2$ separately. Of course, one needs to make
sure that no other competing phase forms, which can be a challenging
task for ternary systems.

\subsection{III. Thermodynamic stability}

{\it Computational details.} In the evaluation of the low temperature
stability of ternary compounds, we rely on the ``{\it ab initio}
formation energy'' criterion, which has been shown to be a reliable
approach for binary systems  \cite{Curtarolo_DMQC,Curtarolo_JMEAS}
(expected probability of predicting the correct experimental compounds
or immiscibilities is $\eta^{\star}\approx97.3\%$ as defined in
equation (5) of Ref.  \cite{Curtarolo_CALPHAD}). {\it Ab initio} total
energy calculations are performed in the generalized gradient
approximation with projector augmented-wave pseudopotentials (PAW)
 \cite{bloechl994} and Perdew, Burke, and Ernzerhof  \cite{PBE}
exchange-correlation functional, as implemented in {\small VASP}
 \cite{kresse1993,kresse1996b}. Because of a significant charge
transfer between metal and boron in most structures considered we use
PAW pseudopotentials in which semicore states are treated as
valence. This is especially important for the Li-B system as discussed
in Refs.  \cite{US_PAW} and  \cite{US1}. Simulations are carried out at
zero temperature and without zero-point motion; spin polarization is
used only for borides of Fe, Co, and Ni. We use an energy cutoff of 398
eV and at least 4000/(number of atoms in unit cell) ${\bf k}$-points
distributed on a Monkhorst-Pack mesh  \cite{MONKHORST_PACK}. All
structures are fully relaxed and numerically converged to within 1-2
meV/atom.

\begin{figure}[t]
  \begin{center}
%    \vspace{-4mm}                                                                                                                                                                
    \centerline{\epsfig{file=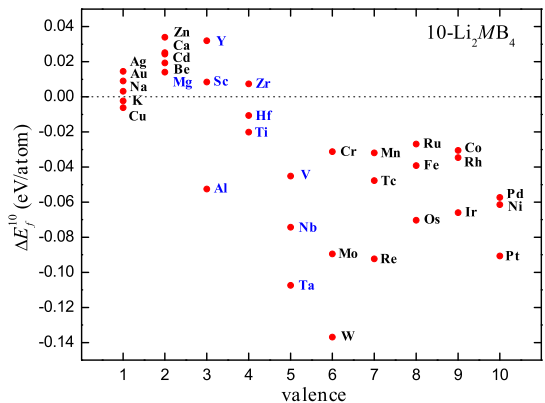,width=80mm,clip=}}
%    \vspace{-8mm}                                                                                                                                                                
    \caption{ \small (Color online).  Relative stability of the                                                                                                                   
      10-Li$_2M$B$_4$ metal borides (Fig. \ref{fig1}(c)) with respect                                                                                                             
      to phase separation into 11-LiB (MS2) and 0-$M$B$_2$ (C32) (see                                                                                                             
      Eq. 1) as a function of metal valence. The group-II-V metals                                                                                                                
      highlighted in blue form stable C32-type diborides.}
    \label{fig2}
  \end{center}
\end{figure}

{\it Stability of Li$_2$$M$B$_4$ phases.} Figure
\ref{fig2} shows the relative formation energy $\Delta E_f^{10}$ of
10-Li$_2M$B$_4$ metal borides (Fig. \ref{fig1}(c)) with respect to
phase separation into 11-LiB (MS2) and 0-MB$_2$ (C32):

\begin{eqnarray} 
\Delta E_f^{10}\equiv E_f^{10}-\frac{4}{7} E_f^{11}-\frac{3}{7} E_f^{0},
\end{eqnarray} 
(all formation energies are per atom). We observe that there are many
metals that can stabilize MS2-LiB and a number of them do have stable
C32-MB$_2$ phases ($M$ = Al, Hf, Ti, V, Nb, Ta)
 \cite{PAULING,US2_note1}. Considering that metals $M$ = Hf, Ti, V, Nb,
Ta have no reported stable compounds with Li or Li-B
 \cite{PAULING,ICSD}, there is a good chance that the predicted layered
phases will form in the Li-$M$-B ($M$ = V, Nb, Ta) ternary
systems. Note that stable structures for several transition-metal
diborides have larger unit cells with buckled boron layers and
different location of metal sites [WB$_2$ ($hP12$, $hR18$), MoB$_2$
(hR18), RuB$_2$ and OsB$_2$ ($oP6$) (the $\delta$ phase
 \cite{US2_note2} is basically a non-corrugated $oP6$)]. For example,
the oP6-RuB$_2$ phase is favored over C32-RuB$_2$ by about 0.4
eV/atom, which would make 10-Li$_2$RuB$_4$ unstable. However, one
could expand the library of possible ternary configurations by
creating more stable MS sequences with buckled boron layers. As for
the mono- and divalent metal diborides, it is not surprising that they
cannot improve the stability of MS-LiB in the corresponding ternary
alloy because they are underdoped themselves  \cite{Oguchi}. Structural
and electronic properties for selected proposed 1010-Li$_2$$M$B$_4$
phases are summarized in Table I.

\begin{table*}[bht]
  \begin{center} 
%  \vspace{-7mm} 
  \caption{ \small Structural and electronic properties of the
  proposed 1010-Li$_2$$M$B$_4$ phases for metals $M$ that have stable
  C32-$M$B$_2$ phases. We list fully relaxed lattice vectors and
  Wyckoff positions for the $P6_3/mmc$ (no. 194) unit cells: B1 ($4f$)
  (1/3,2/3,$z_{B1}$); B2 ($4e$) (0,0,$z_{B2}$); Li ($4f$)
  (1/3,2/3,$z_{Li}$); and $M$ ($2c$) (1/3,2/3,1/4).
  $E^{\sigma,1}_{\Gamma}$ and $E^{\sigma,2}_{\Gamma}$ and positions of
  the boron $\sigma$ states above the Fermi level at the $\Gamma$
  point and $E_f^{1010}$ is a relative formation energy per atom with
  respect to MS2-LiB and C32-$M$B$_2$. Band structures for
  1010-Li$_2$AlB$_4$ and 1010-Li$_2$TiB$_4$ are shown in Figs. 5 and
  6, respectively.}
  \begin{tabular}{c|@{\hspace{7mm}}c@{\hspace{7mm}}c@{\hspace{7mm}}c@{\hspace{7mm}}c@{\hspace{7mm}}c@{\hspace{7mm}}c@{\hspace{7mm}}c@{\hspace{7mm}}c@{\hspace{7mm}}}
    \hline\hline 
    $M$ in            &     $a$    &   $c$      &   $z_{B1}$    &   $z_{B2}$  &   $z_{Li}$    &  $E^{\sigma,1}_{\Gamma}$ & $E^{\sigma,2}_{\Gamma}$ & $E_f^{1010}$ \\
    1010-Li$_2$$M$B$_4$&     (\AA)  &   (\AA)    &               &             &               &     eV          &       eV       &   eV/atom    \\
    \hline 
    Mg &    3.048 &     18.94 &   0.6560 &   0.6560 &   0.4214 &   0.93 &   0.75 &    0.014 \\ 
    Al &    3.012 &     18.46 &   0.6643 &   0.6644 &   0.4198 &   0.62 &   0.14 &   -0.052 \\ 
    Sc &    3.116 &     17.77 &   0.6524 &   0.6526 &   0.4263 &   0.78 &  -0.21 &    0.009 \\ 
    Y  &    3.225 &     16.14 &   0.6298 &   0.6302 &   0.4482 &   0.55 &  -0.13 &    0.033 \\ 
    Ti &    3.040 &     17.99 &   0.6626 &   0.6627 &   0.4197 &   0.28 &  -1.54 &   -0.019 \\ 
    Zr &    3.138 &     17.85 &   0.6514 &   0.6515 &   0.4265 &   0.16 &  -1.25 &    0.008 \\ 
    Hf &    3.119 &     18.20 &   0.6558 &   0.6559 &   0.4229 &   0.05 &  -1.40 &   -0.010 \\ 
    V  &    2.968 &     18.48 &   0.6670 &   0.6670 &   0.4181 &  -0.04 &  -2.43 &   -0.044 \\ 
    Nb &    3.044 &     19.17 &   0.6610 &   0.6610 &   0.4188 &  -0.28 &  -2.45 &   -0.074 \\ 
    Ta &    3.036 &     19.12 &   0.6615 &   0.6615 &   0.4194 &  -0.48 &  -2.61 &   -0.108 \\ 
    \hline\hline 
  \end{tabular} 
  %\vspace{-0.4cm} 
  \label{SUM} 
  \end{center} 
\end{table*} 

{\it Li-Al-B system}. Aluminum is a special case because it is the
least electron-rich metal on the list that provides the desired
stabilization for the ternary MS configurations. The sizeable
50-meV/atom energy gain for 10-Li$_2$AlB$_4$ indicates that this
compound would be stable with respect to C32-AlB$_2$ and the known
off-stoichiometry LiB$_y$ as well  \cite{US2}. For the analysis of the
thermodynamic stability of Li$_{2(x-1)}$Al$_x$B$_2$ we also need to
consider the known stable Li-Al binary phases B32-LiAl,
hR15-Li$_3$Al$_2$, and mS26-Li$_9$Al$_4$  \cite{PAULING,LiAl1,LiAl2}
and the only reported ternary LiAlB$_{14}$ compound  \cite{LiAlB14}
which was observed in experiments on doping AlB$_2$ with
Li \cite{LiAlB1}.

\begin{figure}[t]
  \begin{center}
%    \vspace{-5mm}
    \centerline{\epsfig{file=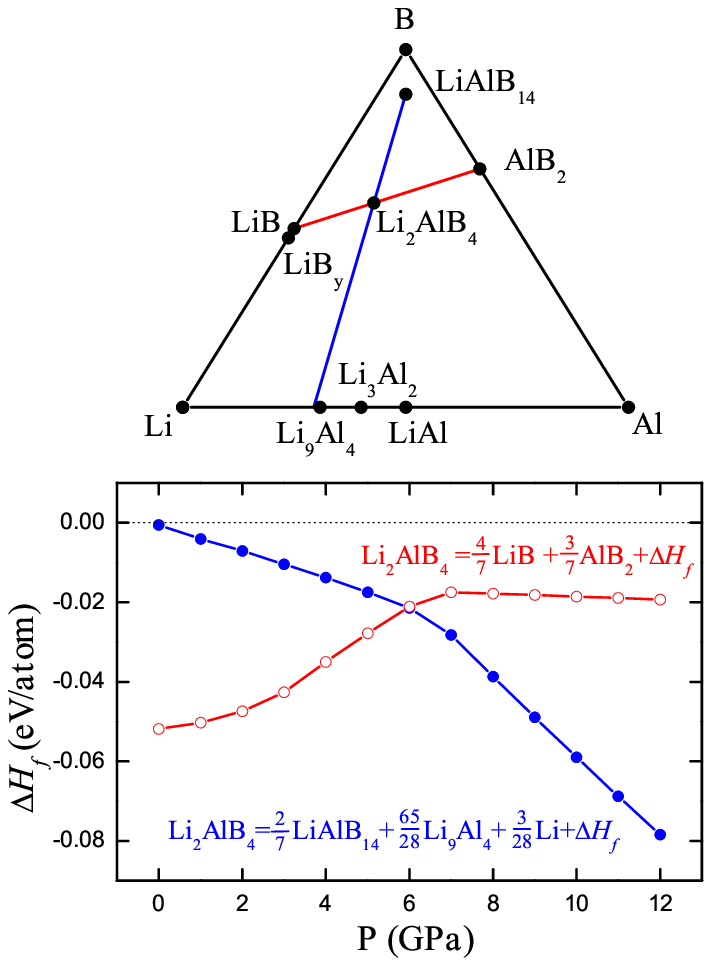,width=80mm,clip=}}
%   \vspace{-8mm}
    \caption{ \small (Color online). Top: location of known and predicted phases on
    the Li-Al-B phase diagram (the boron-rich Li-B and Al-B phases are
    not shown). Bottom: Relative stability of the proposed
    Li$_2$AlB$_4$ with respect to the decomposition along the lines
    shown on the top panel.}
    \label{fig3}
  \end{center}
\end{figure}
%\vspace{-2mm}
%
The location of these phases on the Li-Al-B phase diagram is shown in
Fig. 3. Because 10-Li$_2$AlB$_4$ lies very close to the line
connecting LiAlB$_{14}$ and mS26-Li$_9$Al$_4$, its relative stability
will be defined primarily by the formation energies of these two
compounds. We find that 10-Li$_2$AlB$_4$ is virtually degenerate in
energy with respect to the mixture of LiAlB$_{14}$, mS26-Li$_9$Al$_4$,
and Li at zero temperature and pressure and without zero-point
corrections; the finite temperature contributions must have a negative
effect on the 10-Li$_2$AlB$_4$ relative stability, as there are no
reports on the formation of this compound.

In order to stabilize the proposed phase one could use high pressures:
as we have demonstrated previously the MS phases are unusually
soft \cite{US1,US2}. Indeed, the calculated relative formation enthalpy
as a function of pressure for 10-Li$_2$AlB$_4$ in Fig. 3 either
becomes negative (with respect to LiAlB$_{14}$, mS26-Li$_9$Al$_4$, and
Li) or remains negative (with respect to 11-LiB and 0-AlB$_2$). The
considerable change in slope at about 6 GPa is related to a sudden
$\sim10$\% decrease in the 10-Li$_2$AlB$_4$ atomic volume at that
pressure (Fig. 3). Overall, 10-Li$_2$AlB$_4$ compresses by over 25\%
when the pressure is increased from 0 GPa ($v$ = 10.4 \AA$^{3}$/atom,
$H_f=-0.163$ eV/atom) to 12 GPa ($v$ = 7.7 \AA$^{3}$/atom,
$H_f=-0.302$ eV/atom). For comparison, the LiAlB$_{14}$ phase is much
more compact at $P=0$ GPa ($v$ = 7.7 \AA$^{3}$/atom, $H_f=-0.168$
eV/atom) but compresses by only 5\% at $P=12$ GPa ($v$ = 7.3
\AA$^{3}$/atom, $H_f=-0.217$ eV/atom). These results suggest that
this ternary and other binary boron-rich Li-B and Al-B
phases \cite{PAULING,POLY}, also comprised of rigid boron polyhedra
(e.g., Li$_3$B$_{14}$), should not prohibit the formation of the
predicted layered compounds under the pressures considered.

\begin{figure}[t]
  \begin{center}
%    \vspace{-5mm}
    \centerline{\epsfig{file=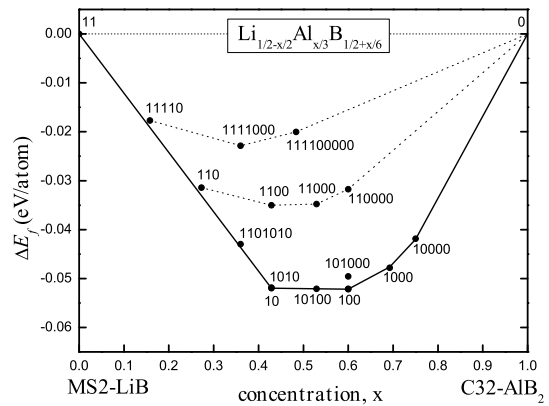,width=80mm,clip=}}
 %   \vspace{-8mm}
    \caption{ \small Relative stability of Li$_{2(1-x)}$Al$_x$B$_2$
      ternary metal borides with respect to the phase separation into
      11-LiB (MS2) and 0-AlB$_2$ (C32). The dotted lines connect
      members of the 110$\cdots$0 and 11110$\cdots$0 series.}
    \label{fig4}
  \end{center}
\end{figure}
%\vspace{-2mm}
%

There is strong evidence that another stable phase has formed in
different experiments involving Li, Al, and
B \cite{LiAlB1,LiAlB2}. Monni {\it et al.} suggested that
their unidentified low-angle spurious peaks could belong to a binary
Li-B phase \cite{LiAlB2}. The authors purposely prepared a sample at the LiB$_2$
composition and indeed observed a matching set of the low-angle
reflections (see Fig. 7 in Ref.  \cite{LiAlB2}). Their conclusion that
the unknown phase is a binary Li-B compound agrees well with the
observation of 12.2$^{\circ}$ and 20.9$^{\circ}$ peaks at 40\% - 50\%
Li compositions by Wang {\it et al.} in
1978 \cite{LiB_Wang78}. However, in a series of detailed studies of the
Li-B system W\"{o}rle {\it et al.} have conclusively determined that
such low-angle peaks correspond to Li$_6$B$_{18}$(Li$_2$O)$_x$ phases
(0$<x<$1) \cite{LiB_B6}. The authors' latest analysis of the X-ray
patterns and electron density distributions has indicated that the
zeolite-like structure of Li$_6$B$_{18}$(Li$_2$O)$_x$ consists of
interconnected boron octahedra with large tunnels filled with neutral
Li$_2$O template \cite{LiB_B6}.
Therefore, it appears that the stable oxygen-containing phase also
formed in the earlier experiments involving Li and B at about 1:2
composition. This knowledge is important for future experiments on
the ternary metal borides: in order to obtain the predicted layered
compounds (whose stability can be affected by a few meV/atom) one
needs to reduce the amount of oxygen in the system.

In order to determine whether there are more stable ternary MS
configurations we calculate a number of Li-Al-B phases with various
stacking sequences and compare their relative stability, $\Delta E_f$,
with respect to the phase separation in 11-LiB and 0-MB$_2$.  The
results in Fig. \ref{fig4} suggest that 10-Li$_2$AlB$_4$ would be the
most LiB-rich phase to form, while one could observe several
AlB$_2$-rich phases at different concentrations (a similar phase
diagram is obtained for the Li-Ti-B system). The effect of permutation
of the ``1'' and ``0'' units on the energy at given composition is
illustrated with the 10$\cdots$0, 110$\cdots$0, 11110$\cdots$0 series
shown in Fig. \ref{fig4}. These results clearly demonstrate the
benefit of having stackings with alternating electron-poor ``1''- and
electron-rich ``0''-units. For instance, 1010 is 20 meV/atom below
1100, which implies that the boron layer in the 11 block of 1100,
surrounded by two Li layers on each side, does not take full advantage
of the available charge provided by Al. This behavior is consistent
with our previous observation that substitutional doping leads to the
filling of boron states only in the layer closest to the dopant
 \cite{US3}.

Note that the boundary of the convex hull
MS2-LiB$\leftrightarrow$C32-AlB$_2$ is determined by the 10$\cdots$0
series, which implies that there is a thermodynamic driving force for
the formation of ``1''-units in the ``0'' matrix. Indeed, doing
extrapolation over the 10$\cdots$0 series we find that replacement of
one Al layer via $n$AlB$_2$ + 2Li = Li$_2$Al$_{n-1}$B$_{2n}$ + Al is
an exothermic reaction with $\Delta E$ = 0.6 eV/(Li atom)
($n\rightarrow\infty$). There is an alternative exothermic reaction
with an even bigger $\Delta E$ = 0.8 eV/(Li atom)
($n\rightarrow\infty$) that preserves the C32 structure: $n$AlB$_2$ +
Li = LiAl$_{n-1}$B$_{2n}$ + Al. Therefore, one would expect to see
substitution of Al layers with Li first, and only at high Li
concentration should the ``1''-units start forming: in the limiting
case of 1:1 Li:Al composition the reaction LiAlB$_4$ + Li =
Li$_2$AlB$_4$ would be energetically favorable by $\Delta E$ = 0.4
eV/(Li atom). However, these reactions have apparently not been
observed due to the formation of the stable ternary LiAlB$_{14}$ and
the oxygen-containing phases \cite{LiAlB2}.

In summary, based on our formation enthalpy calculations,
Li$_2$AlB$_4$ is marginally stable with respect to the considered
known compounds under normal conditions but can be stabilized by
hydrostatic pressure. Considering the stabilizing effect of Al we
expect a lower pressure threshold for the formation of the ternary
MS-Li-Al-B phases compared to the case of the binary MS-Li-B phases.

More detailed studies should be carried out in the future to account
for the finite-temperature contributions in the Gibbs free energy;
this will be a difficult task because the boron-rich metal borides
have large unit cells with fractional occupancies. In addition, one
should consider not only compounds reported for a given ternary
system, but also crystal structure phases observed in similar
systems. For example, the absence of any stable Li-$M$-B phases ($M$ =
V, Nb, Ta) in the ICSD database \cite{ICSD} could simply be an
indication that these ternary systems have not been fully explored
experimentally yet. It would not be surprising then if attempts to
synthesize the predicted layered phases would lead to the formation of
phases not considered here, such as the metal-rich Li$_2$Pd$_3$B
superconductor comprised of linear chains of boron
\cite{Li2Pd3B}. High-throughput simulation of selected ternary systems
identified in our present work is subject of future studies.

\subsection{IV. Electronic properties}

\begin{figure}[t]
  \begin{center}
%    \vspace{-5mm}                                                                                                                                                                
    \centerline{\epsfig{file=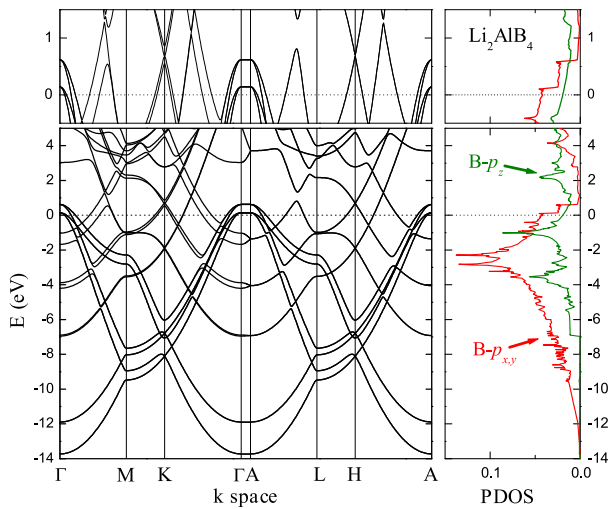,width=80mm,clip=}}
%    \vspace{-8mm}                                                                                                                                                                
    \caption{ \small (Color online). Band structure and                                                                                                                           
      partial density of states (PDOS) in 1010-Li$_2$AlB$_4$. The top                                                                                                              
      panels show the region near the Fermi level (0 eV).}
    \label{fig5}
  \end{center}
\end{figure}

It is interesting to see what changes in properties the combination of
the non-superconducting $M$B$_2$ and potentially superconducting
MS-LiB structures would induce. Electronic properties of the proposed
MS ternary phases are examined for two representative
1010-Li$_2$AlB$_4$ and 1010-Li$_2$TiB$_4$ compounds with hexagonal
unit cells (1010 and 10 phases are nearly degenerate in energy). Boron
layers in the 1010 structure are indistinguishable, but there are two
different sets of degenerate boron states derived from those in the
original ``0'' and ``1'' metal borides. For example, in MS2-LiB the
practically dispersionless $p\sigma$ states of boron are 1.0 eV above
the Fermi level \cite{US1,US2} while in C32-AlB$_2$ they are
completely filled [$E^{\sigma}_{\Gamma}$=-1.6 eV, $E^{\sigma}_A$=-1.0
eV]; in 1010-Li$_2$AlB$_4$ the two sets of boron $p\sigma$ states are
not completely filled, being 0.6 and 0.1 eV above the Fermi
level. This is a very satisfying outcome, because they still
contribute a considerable 0.042
states/(eV$\cdot$spin$\cdot$boron\,atom) to the density of states
(DOS) at the Fermi level (to be compared to 0.059 and 0.049
states/(eV$\cdot$spin$\cdot$boron\,atom) in MS2-LiB and C32-MgB$_2$,
respectively\cite{US2}). Because boron orbitals do not overlap across the
Li-filled portion of 1010-Li$_2$AlB$_4$, the $p\sigma$ states are not
dispersed and give rise to the desired, nearly cylindrical
two-dimensional Fermi surfaces similar to those of MS-LiB
\cite{US1,US2,Mazin}. In addition, the compound has $p\pi$ Fermi
surfaces, as the $p\pi$-bands crossing now happens at about 0.7 eV. The
$p\pi$ states DOS of 0.019 states/(eV$\cdot$spin$\cdot$ boron atom) is
a significant improvement with respect to MS2-LiB, which lacks these
states altogether. However, this contribution is still below the value
in MgB$_2$ (0.032 states/(eV$\cdot$spin$\cdot$ boron atom)) and,
unfortunately, comes at the expense of losing some $p\sigma$ DOS.

\begin{figure}[t]
  \begin{center}
%    \vspace{-5mm}                                                                                                                                                               \
    \centerline{\epsfig{file=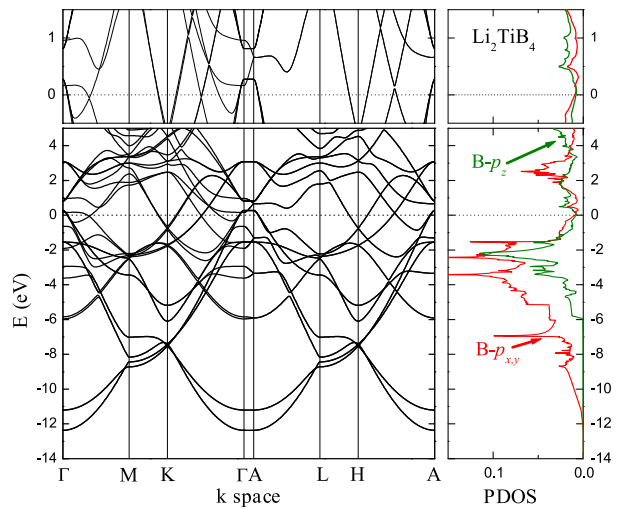,width=80mm,clip=}}
%   \vspace{-8mm}                                                                                                                                                                
    \caption{ \small (Color online). Band structure and partial
      density of states (PDOS) in 1010-Li$_2$TiB$_4$. The top panels
      show the region near the Fermi level (0 eV).}
    \label{fig6}
  \end{center}
\end{figure}

\begin{figure}[htb]
  \begin{center}
%    \vspace{-4mm}
    \centerline{\epsfig{file=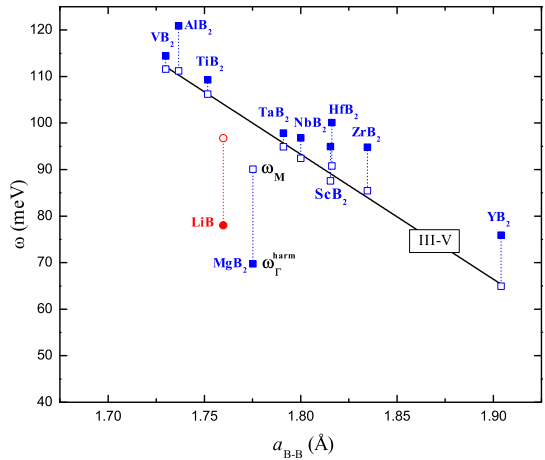,width=80mm,clip=}}
%    \vspace{-8mm}
    \caption{ \small (Color online). Frequency of the in-plane boron
      phonons at M (open symbols) and $\Gamma$ (solid symbols) points
      as a function of boron-boron bond length. The known C32-$M$B$_2$
      diborides are shown as blue squares, the proposed MS2-LiB is
      shown as red circles.  The solid line shows a linear fit of the
      M-point frequencies for the known diborides with metal valence
      from III to V.}
%    \vspace{-6mm}
    \label{fig7}
  \end{center}
\end{figure}
%\vspace{-5mm}
%

\begin{figure}[htb]
  \begin{center}
%    \vspace{-6mm}
    \centerline{\epsfig{file=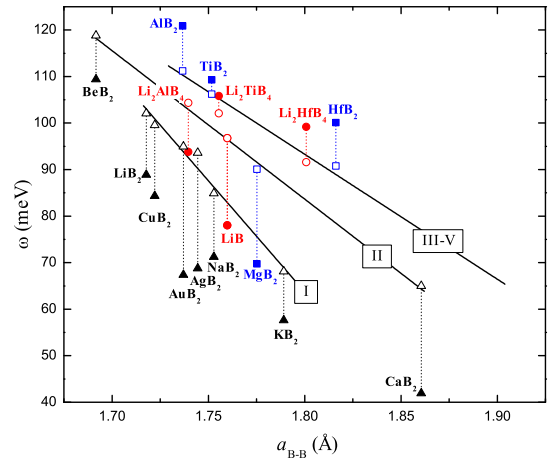,width=80mm,clip=}}
%    \vspace{-8mm}
    \caption{ \small (Color online). Frequency of the in-plane boron
      phonons at M (open symbols) and $\Gamma$ (solid symbols) for
      C32-$M$B$_2$ diborides of group I, II, and III-V metals. The
      blue squares, black triangles, and red circles correspond to the
      known, unstable, and proposed (11-LiB and 1010-Li$_2$$M$B$_4$)
      compounds, respectively. The linear fits to the sets of M point
      frequencies are labeled according to the valence of the metal.}
%     \vspace{-7mm}
   \label{fig8} \end{center}
\end{figure}
%\vspace{-5mm}
%

Figure \ref{fig6} shows that boron in 1010-Li$_2$TiB$_4$ is too
overdoped to have any strong MgB$_2$-type electron-phonon
coupling. The bottom $p\sigma$ band at -1.5 eV is completely filled,
and even though the top band is still 0.3 eV above the Fermi level the
steep $\partial E/\partial k$ derivative on the narrow cylinder along
${\Gamma}$-A results in a relatively low DOS of 0.010
states/(eV$\cdot$spin$\cdot$boron\,atom). 1010-Li$_2$HfB$_4$ exhibits
similar properties but has a slightly higher $p\sigma$ DOS of 0.017
states/(eV$\cdot$spin$\cdot$boron\,atom). 1010-Li$_2$$M$B$_4$
compounds with transition metals of valence V and higher present cases
of borides with completely filled boron $p\sigma$ states (see Table I).

\subsection{V. Phonon softening}

As previously discussed \cite{US3}, the next step in the analysis of
superconducting properties of the proposed materials should be an {\it
  ab initio} calculation of the electron-phonon coupling
$\lambda$. However, the full scale calculations become prohibitively
expensive for compounds with large unit cells, and more so if they
have nested Fermi surfaces. In the following we attempt to obtain
information on the strength of the electron-phonon coupling from the
compounds' vibrational properties.

Indeed, it has been previously argued that the unexpected decrease of
$\lambda$ in MS-LiB with respect to that in MgB$_2$ can be related to
the increase of the in-plane boron frequency and to the smaller
softening of this mode.  \cite{US3,Mazin} The considerations were based
on the fact that for a zone-center optical phonon mode the softening
due to screening by metallic electrons, $\Delta\omega$, is connected
to the Fermi-surface average of the square of the electron-phonon
matrix element, $g$, through $\Delta\omega^2=-4\omega\langle                                                                                                                      
g^2\rangle N(0)$ ($N(0)$ is the electronic DOS at the Fermi level)
 \cite{Mazin,Mazin2,Mazin18}.

Unfortunately, there is no simple way to evaluate the unscreened
frequency in real materials. However, in MgB$_2$-type superconductors
the M-point phonons have wavevectors larger than $2k_F$ and are thus
screened much less than the ${\Gamma}$-point phonons. For this
reason we will use the $\omega_M - \omega_{\Gamma}$ difference to
reveal how the in-plane boron vibrations soften across a set of binary
and ternary layered metal borides. It should be stressed that the
above frequency difference cannot be treated as the real softening for
quantitative estimates of the electron-phonon coupling; nevertheless,
we expect it to be useful for selecting candidates with the strongest
electron-phonon coupling.

For the C32-$M$B$_2$ diborides and MS2-LiB monoboride, we calculate
the frequencies of the in-plane boron mode at ${\Gamma}$ and the TO
mode at M points in the frozen-phonon approach using the fourth-order
corrections and the same settings as described in Refs.
\cite{Kunc,US1}. We neglect the anharmonic and non-adiabatic effects
since we are interested in general trends. Full phonon spectra in
existing metal diborides have been calculated previously
\cite{MgB2_Dispersion_1,MgB2_Dispersion_2}, and the general inverse
dependence of the $E_{2g}$ frequency at ${\Gamma}$ on the boron-boron
bond length has been pointed out \cite{MgB2_Dispersion_2}.  In
Fig. \ref{fig7} we plot, as functions of the boron-boron bond length
$a_{B-B}$, the ${\Gamma}$- and M-point sets of calculated
frequencies. The latter exhibits a much more consistent linear
decrease with the bond length for all known metal diborides of groups
III-V. One can further discern two subsets (Al, Hf, Zr, Y and the
rest) which have ${\Gamma}$-point frequencies following the linear
dependence with similar negative slopes (we see no apparent factor
which could be responsible for this behavior). Note that for the
electron-rich metal diborides $\omega_M$ is below $\omega_{\Gamma}$,
which illustrates why the former cannot be considered as the
unscreened frequency. However, the calculated values of ${\Gamma}$-
and M-point frequencies will be useful as a reference when we examine
the vibrational properties of the ternary MS phases derived from the
corresponding metal diborides.

As expected, MgB$_2$ stands out as a compound with a large softening
of the frequency at ${\Gamma}$. In addition, Fig. \ref{fig7} shows
that the M-point frequency does not follow the general trend of
group-III-V metal diborides either; this behavior cannot be directly
related to the electron-phonon coupling with the $p\sigma$ boron
states. The phenomenon is explained in Figure \ref{fig8} where we plot
calculated frequencies for a large set of group-I-II metal
diborides. The M-point frequency $\omega_M$ for divalent (Be, Mg, Ca)
and monovalent (Li, Cu, Au, Ag, Na, K) metals shows similar and nearly
perfect linear dependences on the boron-boron bond length. One can
expect to have the same trends for the unscreened frequency, only it
would be shifted up by a constant offset.

These results give a new perspective on the relation between the
vibration properties and the strength of the electron-phonon coupling
in LiB discussed in Refs.  \cite{Mazin,US3}. As a boride isovalent to
MgB$_2$, LiB follows the $\omega_M(a_{B-B})$ dependence obtained for
the Be-Mg-Ca series of diborides shown in Fig. \ref{fig8}.
Therefore, one of the key factors in the LiB hardening of the in-plane
frequency [with respect to that in MgB$_2$] is the shortening of the
boron-boron bond length, which depends on the metal and the particular
filling of the $p\pi$ and $p\sigma$ states of boron. Hence, it follows
that the frequency softenings in LiB and MgB$_2$ are, in fact,
comparable.

Finally, we can comment on the superconducting properties of the
proposed ternary compounds. The introduction of electron-rich metals
(group IV and higher) into 1010-Li$_2$$M$B$_4$ overdopes the boron
layers, and causes the vibrational properties to become closer to
those of the corresponding diborides: $\omega_{\Gamma}$ is still above
$\omega_M$ for Ti and Hf (Fig. \ref{fig8}). 1010-Li$_2$AlB$_4$ is a
more promising MgB$_2$-type superconducting material, judging by the
pronounced softening of the in-plane boron mode (note that in AlB$_2$
$\omega_{\Gamma}$ is harder than $\omega_M$). Unfortunately, aluminum
induces the shortening of the bond length and a substantial increase
of the in-plane boron frequency, which will likely weaken the
electron-phonon coupling \cite{Mazin,US3} (the next section is devoted
to a more accurate description of this compound). As for the proposed
MS phases at other compositions, we find that all ternary systems with
negative relative stability for 1010-Li$_2$$M$B$_4$ (Fig. \ref{fig2})
have negative relative stability for 110-Li$_4$$M$B$_6$ as well (the
latter are typically 2-7 meV/atom above the line connecting 11-LiB
and 1010-Li$_2$$M$B$_4$ and are only metastable). Therefore, if
110-Li$_4$$M$B$_6$ could still be synthesized with large in size
transition metals, such as Hf or Ta, the superconducting properties of
the boron layer in the ``11'' portion (which should be well isolated
by the double layers of lithium from the electron-rich ``0''-unit)
would be enhanced due to the stretching of the overall boron bond
length in the compound. A similar phenomenon has been recently
observed in stretched MgB$_2$ films \cite{MgB2_stretched_films}. A
recent Raman study of the vibrational and superconducting properties
of doped MgB$_2$ has revealed that the adjustment of the frequency
alone might not be enough to achieve a higher $T_c$ \cite{Raman_MgB2}.

The trend shown in Figs. \ref{fig7}-\ref{fig8} is corroborated by the
well known correlation between the increase of the electron-phonon
coupling and the inevitable dynamical destabilization of the structure
\cite{Correlation_Testardi,Pickett_review,Correlation_Cordero,Correlation_BaKBiO3_1,Correlation_BaKBiO3_2}
(a too strong renormalization can cause a Peierls-type distortion, a
band Jahn-Teller transformation or a structural transition
\cite{Pickett_review}). In our considered metal boride prototypes, the
specific relation between stability and superconductivity originates
from the subtle effects associated with the filling of binding states
of boron.

The focus of our study has been the MgB$_2$-type
superconductors, while there have been reports on Nb-deficient
superconductors based on C32-NbB$_2$ with $T_c$ of nearly 10
K \cite{MB2,NbxB2}. Indeed, it would be interesting to explore the
superconducting potential of Li$_2$$M$B$_4$ compounds ($M$ =
V,Nb,Ta) which would have a non-MgB$_2$-type electron-phonon
coupling mechanism.

\subsection{VI. Phonon spectrum and electron-phonon coupling.}

In the previous sections we have shown that Li$_2$AlB$_4$ has a
significant amount of boron $p\sigma$ and $p\pi$ DOS at the Fermi
level and exhibits a characteristic softening of the in-plane boron
phonon mode at ${\Gamma}$. All these quantities are lower compared to
those in MgB$_2$ so that one can expect the superconducting properties
of Li$_2$AlB$_4$ to be less appealing. However, a full calculation of
the electron-phonon coupling (for all the modes and all {\bf k}-points
in the Brillouin zone) is needed to understand the superconducting
properties of Li$_2$AlB$_4$. For example, in MgB$_2$ the intercalant
modes are weakly coupled; based on our results for LiB \cite{US3} the
situation could be different in Li$_2$AlB$_4$.  In addition, due to
symmetry constraints in calculations on periodic systems, energy
minimization might not guarantee convergence to an equilibrium free of
dynamical instabilities (imaginary phonon frequency at ${\bf
  q}\ne{\Gamma}$). Thus, dynamical stability has to be checked by
explicit calculation of the phonon frequencies in the whole BZ. In
this section we calculate vibrational phonon frequencies and the
electron-phonon coupling of Li$_2$AlB$_4$ using density functional
theory in the linear response approach \cite{BaroniRMP2001}.

\begin{figure}[htb]
  \begin{center}
        \vspace{-2mm}
    \centerline{\epsfig{file=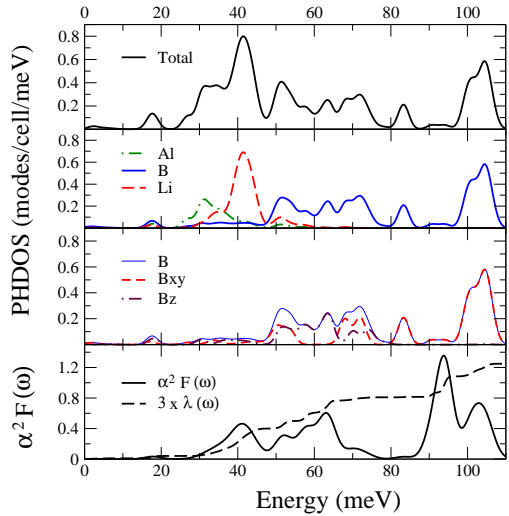,width=80mm,clip=}}
%    \vspace{-5mm}
    \caption{ \small (Color online). Phonon Density of States (PHDOS) decomposed
      over selected vibrations, Eliashberg function $\alpha^2F(\omega)$
      and integrated Eliashberg function $\lambda(\omega)$ in Li$_2$AlB$_4$.}
    \label{fig9}
%   \vspace{-5mm}
  \end{center}
\end{figure}

We find \cite{Phonon_Details} that the theoretically devised structure
of Li$_2$AlB$_4$ is dynamically stable with no imaginary
phonon-frequencies. The obtained phonon density of states (PHDOS) is
plotted in Fig. \ref{fig9}. As it can be seen from the decomposition
along selected cartesian vibrations, B modes are dominant at energies
larger than 50 meV.  In the low energy region ($< 50$ meV), two clear
peaks are seen, one at $\approx 40$ meV due to Li vibrations and the
other one at $30 $ meV due to Al vibrations. As expected from the
large interlayer spacing between the B-layers, B-modes are flat in the
direction perpendicular to the B-layers.

The electron-phonon coupling
$\lambda_{{\bf q}\nu}$ for a phonon mode
$\nu$ with momentum ${\bf q}$ is:
\begin{equation}\label{eq:elph}
  \lambda_{{\bf q}\nu} = \frac{4}{\omega_{{\bf q}\nu}N(0) N_{k}} \sum_{{\bf k},n,m} 
  |g_{{\bf k}n,{\bf k+q}m}^{\nu}|^2 \delta(\epsilon_{{\bf k}n}) \delta(\epsilon_{{\bf k+q}m})
\end{equation}
where the sum is over the Brillouin Zone.
The matrix element is
$g_{{\bf k}n,{\bf k+q}m}^{\nu}= \langle {\bf k}n|\delta V/\delta u_{{\bf q}\nu} |{\bf k+q} m\rangle /\sqrt{2 \omega_{{\bf q}\nu}}$,
where $u_{{\bf q}\nu}$ is the amplitude of the displacement of the phonon, 
 $V$ is the Kohn-Sham potential and $N(0)$ is the electronic density of states 
at the Fermi level.
The calculated average electron-phonon coupling is  
$\lambda=\sum_{{\bf q}\nu} \lambda_{{\bf q}\nu}/N_q\approx 0.41$
($N_{k}$ and $N_{q}$ are the ${\bf k}$-space and ${\bf q}$-space mesh dimensions, respectively \cite{Phonon_Details}).

\vspace{-7mm}

The Eliashberg function
\begin{equation}
\alpha^2F(\omega)=\frac{1}{2 N_q}\sum_{{\bf q}\nu} \lambda_{{\bf q}\nu} \omega_{{\bf q}\nu} \delta(\omega-\omega_{{\bf q}\nu} )
\end{equation}
and the integral $\lambda(\omega)=2 \int_{0}^{\omega} d\omega^{\prime} 
\alpha^2F(\omega^{\prime})/\omega^{\prime}$ 
are shown in Fig. \ref{fig9}.
As can be seen, the contribution of the coupling due to in-plane vibration is 
substantially reduced with respect to MgB$_2$. This confirm what could be qualitatively
inferred from the softening calculated in  Fig. \ref{fig8}.
On the contrary the coupling to intercalant modes is not negligible.

The critical superconducting temperature is estimated using the McMillan
formula  \cite{mcmillan}:
\begin{equation}
T_c = \frac{\langle \omega \rangle_{log}}{1.2}\, \exp \left[{ - \frac{1.04 (1+\lambda)}{\lambda-\mu^* (1+0.62\lambda)}}\right]
\label{eq:mcmillan}
\end{equation}
where $\mu^*$ is the screened Coulomb pseudopotential and
\begin{equation} 
\langle\omega\rangle_{log} = e^{\frac{2}{\lambda}\int_{0}^{+\infty} 
\alpha^2F(\omega)\log(\omega)/\omega\,d\omega }
\end{equation}
the phonon frequencies logarithmic average. We obtain $\langle\omega\rangle_{log}=59.4$ meV
leading to $T_c$ of approximately 3.6 K for $\mu^{*}=0.1$. 
This value might be enhanced by multiband effects.

\subsection{VII. Summary}

We have demonstrated that there is a number of Li-$M$-B systems, in
which MS ternary borides gain in stability with respect to MS2-LiB and
C32-$M$B$_2$. The finding suggests that these potential
superconductors could be synthesized under normal conditions or grown
on metal diboride substrates with a matching lattice constant. We show
that some boron states in the MS ternary compounds [derived from those
in MS2-LiB and C32-$M$B$_2$] are still hole-doped and could give rise
to the MgB$_2$-type superconductivity. In order to pre-select
compounds with the strongest electron-phonon coupling we examine the
softening of the in-plane boron phonon mode in a large class of metal
borides. We find a very well defined correlation between the frequency
of this mode and both the boron-boron bond length and the valence of
the metal. This observation allows us to give a new interpretation of
the hardening of the in-plane boron phonon mode in MS-LiB and identify
one compound, Li$_2$AlB$_4$, that should be a superconductor with
$T_c$ of about 3.6 K. 

We thank R. Margine, L. Boeri, W. Setyawan, M. Mehl, A. Liu and
I. Mazin for valuable discussions. We acknowledge the Teragrid-TACC
center, the Pennsylvania State supercomputer center, and the IDRIS
supercomputing center (project 081202) for computational
support. Research supported by ONR (N00014-07-1-0878) and NSF
(DMR-0639822).


\vspace{-3mm}
\begin{thebibliography}{99}
\vspace{-3mm}

\bibitem{origin}
J. Nagamatsu, N. Nakagawa, T. Muranaka, Y. Zenitani, and J. Akimitsu,
Nature {\bf 410}, 63 (2001).

\bibitem{band_filling}
J. Kortus, O. V. Dolgov, R. K. Kremer, and A. A. Golubov,
Phys. Rev. Lett. {\bf 94} 027002 (2005) and references therein.

\bibitem{CaAlSi}
%Superconducting properties of single-crystalline Ca(Al-0.5,Si-0.5)(2): A ternary silicide with the AlB2-type structure
M. Imai, E. Abe, J. Ye, K. Nishida, T. Kimura, K. Honma, H.
Abe, and H. Kitazawa, Phys. Rev. Lett. {\bf 87}, 077003 (2001).

\bibitem{TaB2} 
H. Rosner, W. E. Pickett, S.-L. Drechsler, A. Handstein, G. Behr,
G. Fuchs, K. Nenkov, K.-H. Muller, and H. Eschrig, PRB {\bf 64},
144516 (2001)

\bibitem{MB2}
A. Yamamoto, C. Takao, T. Masui, M. Izumi, and S. Tajima, Physica C {\bf 383}, 197 (2002).

\bibitem{NbxB2}
Z.-A. Ren, S. Kuroiwa, Y. Tomita and J. Akimitsu, Physica C: Superconductivity {\bf 468}, 411 (2008)

\bibitem{LiBC}
H. Rosner, A. Kitaigorodsky, and W.E. Pickett, Phys. Rev. Lett. {\bf 88}, 127001 (2002).

\bibitem{LixBC}
A. M. Fogg, J. B. Claridge, G. R. Darling, and M. J. Rosseinsky, Chem. Commun. (Cambridge) {\bf 12} 1348 (2003);
A. M. Fogg, J. Meldrum, G. R. Darling, J. B. Claridge, and M. J. Rosseinsky, J. Am. Chem. Soc. {\bf 128}, 10043 (2006).

\bibitem{SCGIC1}
T. E. Weller, M. Ellerby, S. S. Saxena, R. P. Smith, and N. T. Skipper, Nature Phys. {\bf 1}, 39 (2005). 

\bibitem{SCGIC2}
N. Emery, C. Herold, M. d'Astuto, V. Garcia, C. Bellin, J. F. Mareche, P. Lagrange, and G. Loupias, 
Phys. Rev. Lett. {\bf 95}, 087003 (2005). 

\bibitem{SCGIC3}
A. Gauzzi, {\it et al.}, Phys. Rev. Lett. {\bf 98}, 067002 (2007) 

\bibitem{US1}
A. N. Kolmogorov and S. Curtarolo, Phys. Rev. B {\bf 73}, 180501(R) (2006).

\bibitem{MgC}
P. Zhang, S. Saito, S. G. Louie, and M. L. Cohen, Phys. Rev. B {\bf 77}, 052501 (2008).

\bibitem{ICSD}
Inorganic crystal structure database, http://icsd.ill.fr/icsd/index.html

\bibitem{PAULING}
P. Villars, K. Cenzual, J. L. C. Daams, F. Hulliger, T. B. Massalski, H. Okamoto, K. Osaki, A. Prince, and S. Iwata,
Crystal Impact, {\it Pauling File. Inorganic Materials Database and Design System},
Binaries Edition, ASM International, Metal Park, OH (2003).

\bibitem{Oguchi}
T. Oguchi, J. Phys. Soc. Jpn. {\bf 71}, 1495 (2002).

\bibitem{dope_review} 
R.J. Cava, H.W. Zandbergen, and K. Inumaru, Physica C {\bf 385}, 8 (2003). 

\bibitem{MgB2_defects}
F. Bernardini and S. Massidda, Europhys. Lett. {\bf 76}, 491 (2006).

\bibitem{MgAlLiB1}
F. Bernardini and S. Massidda,
Phys. Rev. B {\bf 74}, 014513 (2006). 

%\bibitem{MgBLiC}
%{\sf we should put this one, see email, if you agree let`s find the real citation}
%http://arxiv.org/abs/0704.3526

\bibitem{Palnichenko}
A. V. Palnichenko, O. M. Vyaselev, and N. S. Sidorov, JETF Lett. {\bf 86}, 272 (2007).

\bibitem{Chepulskyy}
R. V. Chepulskii, I. I. Mazin, and S. Curtarolo, 
{\it First-principles search for potential high temperature superconductors 
in the Mg-B-A (A=alkali and alkaline earth metals) system with high boron content}, (2008).

\bibitem{US2}
A. N. Kolmogorov and S. Curtarolo, 
Phys. Rev. B {\bf 74}, 224507 (2006). 

\bibitem{Mazin}
A. Y. Liu and I. I. Mazin, 
Phys. Rev. B {\bf 75}, 064510 (2007). 

\bibitem{US3}
M. Calandra, A. N. Kolmogorov, and S. Curtarolo, 
Phys. Rev. B {\bf 75}, 144506 (2007). 

\bibitem{US4}
A. N. Kolmogorov, R. Drautz, D. G. Pettifor, Phys. Rev. B {\bf 76}, 184102 (2007),

\bibitem{MgAlB4}
J. Q. Li, L. Li, F. M. Liu, C. Dong, J. Y. Xiang, and Z. X. Zhao,
Phys. Rev. B {\bf 65}, 132505 (2002). 

\bibitem{Curtarolo_DMQC}
S. Curtarolo, D. Morgan, K. Persson, J. Rodgers, and G. Ceder, Phys. Rev. Lett. {\bf 91}, 135503 (2003). 

\bibitem{Curtarolo_JMEAS}
D. Morgan, G. Ceder, and S. Curtarolo, Meas. Sci. Technol. {\bf 16}, 296-301 (2005). 

\bibitem{Curtarolo_CALPHAD}
S. Curtarolo, D. Morgan, and G. Ceder, Calphad {\bf 29}, 163-211 (2005). 

\bibitem{bloechl994}
P. E. Blochl, 
Phys. Rev. B {\bf 50}, 17953 (1994). 

\bibitem{PBE}
J. P. Perdew, K. Burke, and M. Ernzerhof, 
Phys. Rev. Lett. {\bf 77} 3865 (1996). 

\bibitem{kresse1993}
G. Kresse and J. Hafner, Phys. Rev. B {\bf 47}, 558 (1993). 

\bibitem{kresse1996b}
G. Kresse and J. Furthmuller, 
Phys. Rev. B {\bf 54}, 11169 (1996). 

\bibitem{US_PAW}
G. Kresse and D. Joubert, 
Phys. Rev. B {\bf 59}, 1758 (1999). 

\bibitem{MONKHORST_PACK}
J. D. Pack and H. J. Monkhorst, {Phys. Rev. B} {\bf 13}, 5188 (1976);
{\bf 16}, 1748 (1977). 

\bibitem{US2_note1}
Figure 3 of Ref  \cite{US2}.

\bibitem{US2_note2}
Figure 2 of Ref  \cite{US2}.

\bibitem{LiAl1}
M. Sluiter, D. de Fontaine, X. Q. Guo, R. Podloucky, and A. J. Freeman, 
Phys. Rev. B {\bf 42}, 10460 (1990). 

\bibitem{LiAl2} 
M.H.F. Sluiter, Y. Watanabe, D. de Fontaine, and Y. Kawazoe,
Phys. Rev. B {\bf 53}, 6137 (1996).  

\bibitem{LiAlB14}
T. Ito and I. Higashi, Acta Cryst. B {\bf 39}, 239 (1983).

\bibitem{LiAlB1}
T. Sun {\it et al. }
J. Superconductivity {\bf 17}, 473 (2004). 

\bibitem{POLY}
Consideration of these binary Li-B and Al-B compounds with large unit
cells and fractional occupancies of atomic sites is beyond the scope
of the present study.

\bibitem{LiAlB2} 
M. Monni {\it et al. }
Phys. Rev. B {\bf 73}, 214508 (2006). 

\bibitem{LiB_Wang78}
F.E. Wang, M.A. Mitchell, R.A. Sutula, and J.R. Holden, J. Less-Common Metals {\bf 61}, 237 (1978).

\bibitem{LiB_B6}
M. W\"{o}rle, R. Nesper, G. Mair, and H.G. von Schnering, Solid State Science {\bf 9}, 459 (2007) and references therein.

\bibitem{Li2Pd3B}
  %Superconductivity in the metal rich Li-Pd-B ternary boride 
K. Togano, P. Badica, Y. Nakamori, S. Orimo, H. Takeya, and K. Hirata, Phys. Rev. Lett. {\bf 93}, 247004 (2004).

\bibitem{Mazin2}
I. I. Mazin and V. P. Antropov, 
Physica C {\bf 385}, 49 (2003).

\bibitem{Mazin18}
C. O. Rodriguez {\it et al. }
Phys. Rev. B {\bf 42}, 2692 (1990).

\bibitem{Kunc}
K. Kunc, I. Loa, K. Syassen, R. K. Kremer, and K. Ahn, J. Phys.: Condens. Matt. {\bf 13}, 9945-9962 (2001). 

\bibitem{MgB2_Dispersion_1}
K.-P. Bohnen, R. Heid, and B. Renker, Phys. Rev. Lett. {\bf 86}, 5771 (2001). 

\bibitem{MgB2_Dispersion_2}
R. Heid, B. Renker, H. Schober, P. Adelmann, D. Ernst, and K. -P. Bohnen, Phys. Rev. B {\bf 67}, 180510(R) (2003). 

\bibitem{MgB2_stretched_films}
A. V. Pogrebnyakov {\it et al. }
Phys. Rev. Lett. {\bf 93}, 147006 (2004). 

\bibitem{Raman_MgB2}
%Raman study of element doping effects on the superconductivity of MgB2
 W. X. Li, Y. Li, R. H. Chen, R. Zeng, S. X. Dou, M. Y. Zhu, and H. M. Jin,
 Phys. Rev. B {\bf 77}, 094517 (2008).

\bibitem{Correlation_Testardi}
  L.R. Testardi,
  in {\it Physical Acoustics}, edited by W.P. Mason and R.N. Thurston (Academic, New York, 1973).

\bibitem{Pickett_review}
  W.E. Pickett,
  % The next breakthrough in phonon-mediated superconductivity
  Physica C {\bf 468}. 126 (2008).

\bibitem{Correlation_Cordero}
  F. Cordero, R. Cantelli, G. Giunchi and S. Ceresara
  %Search for incipient lattice instabilities in MgB2 by anelastic spectroscopy
  Phys. Rev. B {\bf 64}, 132503 (2001).

\bibitem{Correlation_BaKBiO3_1}
  S. Zherlitsyn {\it et al. }
  %, , B. L<FC>thi, V. Gusakov, B. Wolf, F. Ritter, D. Wichert, S. Barilo, S. Shiryaev, C. Escribe-Filippini, and J.L. Tholence,
  %Structural instability and superconductivity in Ba/sub 1-x/K/sub x/BiO/sub 3/
  Eur. Phys. J. B {\bf 16}, 59 (2000).

\bibitem{Correlation_BaKBiO3_2}
  M. Braden {\it et al. }
  %, ,, W. Reichardt, E. Elkaim, J.P. Lauriat, S. Shiryaev, and S.N. Barilo, 
  % Structural distortion in superconducting Ba1-xKxBiO3
  Phys. Rev. B {\bf 62}, 6708 (2000).

\bibitem{BaroniRMP2001}
S. Baroni, S. de Gironcoli, A. Dal Corso, and P. Giannozzi, 
%{\it Phonons and related crystal properties from density-functional perturbation theory},
Rev. Mod. Phys. {\bf 73}, 515 (2001).

\bibitem{Phonon_Details} 
We use the Quantum-espresso code (www.pwscf.org), 
norm-conserving pseudopotentials, the generalized gradient approximation (PBE) and a 60 Rydberg cutoff for the
kinetic energy. We perform the electronic integration using 
$8\times8\times8$ {\bf k}-point mesh. The dynamical matrices were computed
on a $N_q=4\times4\times4$ {\bf q}-point mesh centered at ${\Gamma}$. The electron-phonon 
coupling was computed using a $N_k=30\times30\times30$ {\bf k}-point mesh.

\bibitem{mcmillan} 
  W. L. McMillan, Phys. Rev. {\bf 167}, 331 (1968).
  
\end{thebibliography}
\end{document}